\documentclass[pre,floats,superscriptaddress]{revtex4}

\usepackage{graphicx}
\usepackage{epstopdf}
\pdfoutput=1
\usepackage{amssymb}
\usepackage{amsmath,bm}
\usepackage{psfrag}
\usepackage{epsfig}
\usepackage{float}

\begin{document}
\title{Height transitions, shape evolution, and coarsening of equilibrating quantum nanoislands}

\author{Mikhail Khenner}
\affiliation{Department of Mathematics and Applied Physics Institute, Western Kentucky University, Bowling Green, KY 42101}

\begin{abstract}

Morphology evolution and coarsening of metal nanoislands 
is computed 
within the framework of a surface diffusion-type model that includes the effects of the electron energy confinement within the film, the charge spillage at the film/substrate interface, 
the energy anisotropy of the film surface and the surface stress. The conditions that result in large
islands with flat tops, steep edges, and strongly preferred heights are determined. A strong influence of the film height on the coarsening dynamics and final morphologies is found; the conditions leading to interrupted coarsening are highlighted. The dependence of the geometric parameters of the equilibrium island on the film height and on the island initial volume is computed.
\vspace{0.2cm}\\
\textit{Keywords:}\ Heteroepitaxial ultra-thin metal films, quantum size effect, surface diffusion, coarsening, solid-on-solid wetting
\end{abstract}

\date{\today}
\maketitle


\section{Introduction}
\label{Intro}

Formation of a nanoscale islands on the surfaces of an ultrathin metal films was studied extensively, especially since it was discovered that a low temperature ($<$140K)
deposition of a nonwetting Ag film on GaAs(110), followed by prolonged annealing at moderate temperature (300-400K), results in the atomically flat film above a certain critical film height (thickness) \cite{Smith, Yu}.
Soon after this discovery it was shown \cite{Zhang} that the quantum size effect (QSE), which is rooted in the quantization of the electronic energy in the direction across the 
metal film, is mainly responsible for this new ``electronic" or ``quantum" growth mode. 
Subsequent exploration of the quantum growth in heteroepitaxial Ag/Si(111) and Pb/Si(111) systems ``revealed the presence of large
islands with flat tops, steep edges, and strongly preferred heights". (Quoted from Ref. \cite{Ozer}; 
also see the review \cite{Han1} and references therein.) Resolved Scanning Tunneling Microscopy (STM) images of a 3D island morphologies can be seen, for instance, 
in Fig. 2 of Ref. \cite{Ozer} and in Figures 2 and 17 of Ref. \cite{Han1}. A detailed understanding of the kinetic pathways to form such ``magic"  island morphologies remains incomplete \cite{Ozer,Han1,UFTE,HUQJJLTE,YJEWWNZS}.

Different models were proposed \cite{UFTE,HUQJJLTE,Li,Kuntova} to help explain the experimental findings in some metal/substrate systems (Pb/Si, Ag/AlPdMn, and Ag/NiAl, respectively). 
Ref. \cite{Li} presents the rate equation-type model. The islands are approximated as circular mesas with radii $R_i$ and heights $h_i$, and the evolution equation for the number of atoms in the $i$-th island is constructed and solved numerically, assuming the diffusion of adatoms on the substrate. The QSE is incorporated through a term in the island chemical potential, in a fashion very similar to this paper.
That model, despite its apparent success at predicting the time-evolution of quantities such as the area fraction of the islands of a particular height, is not capable of tracking
neither the surface morphology as a whole, nor the morphology of the individual islands. Besides, since it is limited by its starting assumption of the circular islands, it is unclear how the deviations
from this shape would affect the model results. In Ref. \cite{UFTE} a step dynamics model is developed of the growth of a 
single multilayer island by incorporation of atoms deposited within its capture zone. The island is assumed to have a ``wedding cake" morphology, i.e. $i$ circular layers with 
the radii $r_i(t)$. The conditions for creation of new top layers are prescribed, and the evolution of the radii are determined by the net attachment fluxes of diffusing adatoms to
the edges of the circular layers. The fluxes, in turn, are obtained by solving the appropriate deposition-diffusion equations for adatom density on each terrace. The QSE enters
rather indirectly through prescribing the enhanced adsorption energy on the top of the second layer, with the result that the growth of the 3rd layer is enhanced and the growth of the
4th layer is inhibited. The model thus emphasizes the growth of three-layer circular islands, with such height selectivity indeed dominating in the Ag/AlPdMn system. In Ref. 
\cite{HUQJJLTE} the Kinetic Monte Carlo simulations are performed of the atomistic lattice-gas model of film growth. The Density Functional Theory (DFT) calculations provide adatom
adsorption energies (which reflect QSE promoting a bilayer island formation), interaction energies and key diffusion barriers. Lastly, Monte Carlo simulations of the post-deposition evolution of an
ensemble of Pb islands are presented in Ref. \cite{Kuntova}. Here, the $(i,j)$ island is assumed in the form of a rectangle of height $h_i$ and width $j$. The growth is driven by 
the competition of probabilities $\pi_1$ and $\pi_2$, where the former probability is one of an atom jump from a wetting layer onto island's sidewall (and thus the corresponding
flux increases the island's width), and the
Arrhenius probability $\pi_2$ is one of an atom jump from a sidewall onto the top of the island (thus the corresponding
flux increases the island's height). 
The cited models of the islands ensembles \cite{Li,Kuntova} are most closely related to the study in this paper. However, these models do not allow studies of the dynamics of the morphology evolution and the islands coarsening, 
since the individual island shape is pre-determined and is a constant of the model (i.e., a rectangle). Nonetheless, these models (see also Refs. \cite{Kuntova1,Kuntova2}) are quite successful in predicting the 
equilibrium distributions of the islands heights and widths as functions of the parameters, such as various diffusion and attachment barriers on the island's 
sidewalls and the top surface.
Other models
are even more restrictive by the assumption of a single island of a pre-determined shape \cite{UFTE,HUQJJLTE}. 

In order to enable the studies of the morphology evolution of a metal film surface without the above-mentioned constraints, 
the author recently introduced a new model \cite{JEM_Khenner} that may be seen as partially complementary to Ref. \cite{Li,Kuntova} The new model is a specification to material systems with strong QSE of the classical surface diffusion model 
that was very successfully used to study the morphology evolution in a 
semiconductor-on-semiconductor systems \cite{SDV,Chiu,GolovinPRB2004,Korzec,Korzec1,KTL,LLRV,Aqua}. 
Such specification is primarily achieved by using the form of the film surface energy that very roughly mimics its oscillation 
with the film height due to QSE in a real metal/substrate system \cite{Hirayama,Han,Wu,OPL}. Precisely, the oscillation of the surface energy is taken as simply periodic 
and with a constant period. This assumption does not hold for many ultrathin metal films (Pb is the most extensively studied case), since in such films the surface energy often displays a complicated multi-layer periodicity
(sometimes with a superimposed beating pattern) \cite{Zhang,Han1,Wu}.
Thus the model at this stage is generic and phenomenological. Still, the numerical studies 
of the (nonlinear) islands dynamics that we present in this paper reveal many minutiae details of the islands formation, equilibration, and coarsening that 
are seen in experiment \cite{Smith,Ozer,UFTE,HUQJJLTE,Han1,Hirayama,CHPPM,YJEWWNZS,Li}. 

It is important that
the new model predicts the final quantized island heights and associated island morphologies based (at large) on the form of the dominant QSE 
surface energy that \emph{directly} follows from the quantum theories \cite{OPL,Hirayama,Han}, with the physical parameters in this 
expression (Eq. (\ref{gammaQSE}) below) that have clear physical meaning, and without any fitting or otherwise completely or partially ad-hoc treatment. Some other models that we are aware of \cite{UFTE,HUQJJLTE,Li} assume, tacitly or explicitly, that the diffusion along the island sidewalls is negligible in comparison with the peripheral diffusion along the edge of the island. 
This assumption helps in obtaining (or preserving) certain ``magic" island heights that are detected in the experiment. The new model does not employ this assumption or similar assumptions, thus 
the ``magic" island heights emerge spontaneously through the standard surface diffusion-driven dynamics - given only the initial surface morphology 
(which often does not manifest islands, rather it is composed of bumps and pits). We caution however, that the numerical values of these heights that we calculate
and compute in this paper cannot be taken literally as the approximations to their counterparts in a real system (due to the above-mentioned approximate nature of the QSE oscillation employed by the model and values of most parameters that are at best typical for most metal films). 
In fact, the ``magic" island heights measured in experiment are the integer-multiples of a monolayer (ML) height, while such heights in this paper 
are estimated as the non-integer numbers of monolayers. Nonetheless, the equilibration and coarsening pathways that we compute and that lead to these heights are generally valid
and persist in any quantum system, irrespective of its particular height-selection protocol imposed by a special QSE oscillation and the material parameters.

Another remark is that the annealing temperature of $<$400K is sufficiently low to 
preserve QSE as the major effect \cite{Golubovic}, and also sufficiently high to provide the necessary adatom mobility for the morphology evolution on a very large time scale of 24 hours or longer. In fact, in Pb films on Ni(111) the pronounced QSE was observed at even higher temperatures (up to 474K) \cite{BGZP}. This justifies the use of a continuum model.
A caveat is that it is universally accepted that continuum models are useful for thick films (i.e., with heights $h>\sim 50$ ML); it is less known and appreciated that such models are very successful in the 
description of the dynamics of very thin (liquid or solid) films (the dynamics in either case is governed by a fourth-order PDE). For example, in Ref. \cite{Becker}
the dewetting of a sub-4nm thick liquid film is captured by a thin-film model (which, it is worth mentioning, is designed using the lubrication approximation; this is the analog of
thin-film scaling employed in this paper), and in Ref. \cite{Golubovic} the extended Mullins-type model is used for solid films with heights as small as 1 ML to model cluster growth
during dewetting. More generally, in the mature field of solid-state dewetting the continuum models based on the Mullins-type approach deliver results that are in the semi-quantitative agreement with
experiment \cite{Cheynis,Srolovitz}; these models typically consider a retracting edge of the film, with one of the boundary conditions $h=0$ there. 
Other examples include modeling of films rupture \cite{Vaynblat}. Thus even the infinitesimally small film 
thickness is not, per se, a factor that invalidates the continuum model approach.

To frame our study in the context of current and future applications, we notice that self-organized metal nanoisland arrays are important for emerging technologies based on nonlinear optics, 
plasmonics, photovoltaics, and photocatalysis. Such arrays hold large promise in the design of spectrally selective absorbers \cite{Held},
enhanced fluorescence and infrared spectroscopic devices \cite{Aslan,Jensen}, polarizers and spectral filters \cite{Heger}, and molecular detectors and biosensors \cite{Anker, Liao}.
They are also of interest for manufacture of the next-generation solar cells \cite{Santbergen}. Our modeling study is a step toward design of the film growth protocols that would result in regular arrays that are in demand by these applications.

\section{The model for surface morphology evolution}
\label{Formulation}

In this section we briefly describe the model that enables the computation of the post-deposition dynamics of metal nanoislands. Also we provide the summary of key results of the linear stability analysis (LSA) for
the trivial equilibrium (the planar film surface); for more details, see Ref. \cite{JEM_Khenner}.

The model is based on the Mullins' surface diffusion equation: 
\begin{equation}
h_t=\sqrt{1+h_x^2+h_y^2}\ \mathcal{D}\nabla_s^2 \mu,
\label{MullinsEq}
\end{equation}
where $h(x,y)$ is the height of the surface of  the thin film above the substrate, $x,\;y$ the coordinates in the 
substrate plane, $z$ be the coordinate normal to the substrate plane, $\nabla_s^2$ 
the surface Laplacian, $\mathcal{D}=\Omega^2 D \nu/k T$ the diffusion constant ($\Omega$ is the adatom volume, $D$ the adatom diffusivity, 
$\nu$ the surface density of the adatoms, $k T$ the Boltzmann's factor),
and $\mu=\frac{\delta}{\delta h}\int \gamma\; dS$ the chemical potential;
here $\frac{\delta}{\delta h}$ is the functional derivative, the integral is over the film surface, and
\begin{equation}
\gamma = \gamma_0 + 
\gamma^{(QSE)} + \gamma^{(S)} +\gamma^{(Anis)}
\label{totalEnergy}
\end{equation}
is the surface energy. In Eq. (\ref{totalEnergy}) 
$\gamma_0$ is the nominal (constant) surface energy, and the contributions labeled ``QSE", ``S" and ``Anis" are due to quantum size effect, the stress and the anisotropy, respectively. 

Due to crystalline anisotropy, the surface energy depends on the surface orientation ${\bf n}$, and QSE and the stress introduce a dependence on $h$, i.e. $\gamma = \gamma({\bf n},h)$.
Then the calculation of the functional derivative leads to a four-term expression \cite{Korzec} (see also Refs. \cite{Korzec1,GolovinPRB2004,KTL})
\begin{equation}
\mu = \mu_\kappa+\mu_{wet}+\mu_{anis}+\mu_{h.o.t.},
\label{mu}
\end{equation}
with
\begin{eqnarray}
\mu_\kappa &=& \gamma \kappa,\quad \mu_{wet} = n_3 \partial_h \gamma, \label{mu_kappa_mu_wet}\\
\mu_{anis} &=& -2n_3\left[\left(h_x h_{xx}+h_yh_{xy}\right)\partial_{h_x}\gamma+\left(h_y h_{yy}+h_xh_{xy}\right)\partial_{h_y}\gamma\right]-\frac{\partial_x\partial_{h_x}\gamma+
\partial_y\partial_{h_y}\gamma}{n_3}, \label{mu_anis}\\
\mu_{h.o.t.} &=& \partial_{xx}\partial_{h_{xx}}\frac{\gamma}{n_3}+\partial_{xx}\partial_{h_{yy}}\frac{\gamma}{n_3}+
\partial_{yy}\partial_{h_{xx}}\frac{\gamma}{n_3}+\partial_{yy}\partial_{h_{yy}}\frac{\gamma}{n_3},
\label{mu_hot}
\end{eqnarray}
where $\kappa$ is the curvature, $n_3$ the $z$-component of ${\bf n}$, and ``wet", ``anis", and ``h.o.t." are the short-hand notations for wetting, anisotropy and higher order terms, respectively.

\begin{figure}[h]
\centering
\includegraphics[width=6.0in]{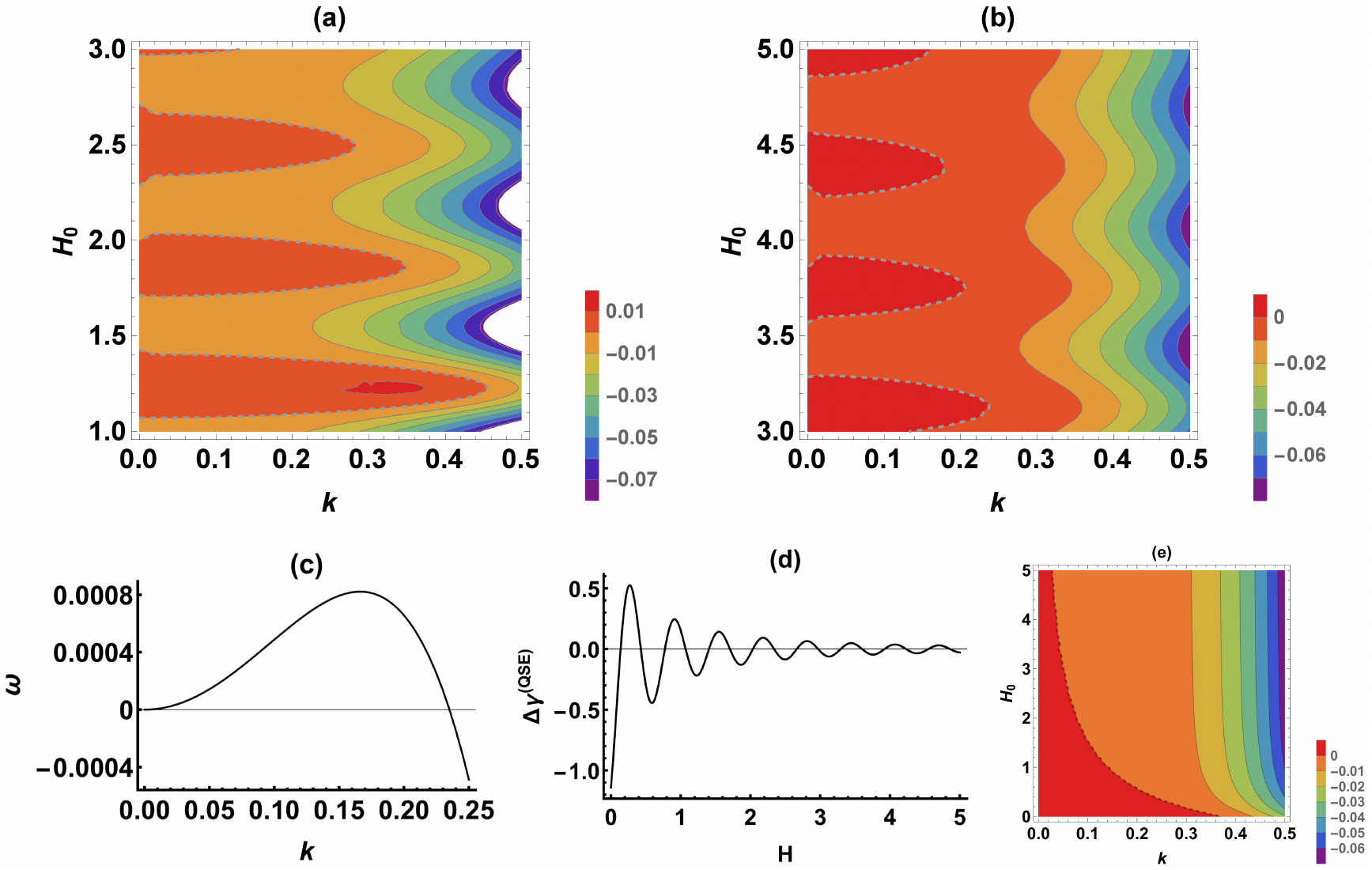}
\vspace{-0.15cm}
\caption{(Color online.) (a,b): Contour plots of the perturbation growth rate $\omega\left(k,H_0\right)$. Dashed lines mark the $\omega=0$ contour. 
From these contour plots (and from the similar ones at larger $H_0$), the stable heights are $H_0=1.5, 2.1, 2.8, 3.4, 4.1, 4.7, 5.3, 6.0, 6.6, 7.2, 7.8, 8.4, 9.1, 9.7, 10.4, 11.0, 11.7, 12.3$.
Each value in this set is, very approximately, the midpoint of an enclosing interval of the stable heights; these values (actually used in the computations described in Sec. \ref{EvolveIslands}) are the 
approximations of the $H$-coordinates of the points of local maximum of the $\Delta\gamma^{(QSE)}(H)$ curve in panel (d). The width of the stable intervals is approximately 0.25, which translates into 0.3 ML
(given the chosen length scale $s = 0.4nm \sim 1.2$ ML). The separation between the stable heights in the above set is $0.6 \div 0.7 \sim 0.8$ ML, and the first
value in the set $H_0=1.5 \sim 1.8$ ML. (All ML values are estimated for Pb.) (c): the cross-section of the plot in (b) by the 
line $H_0=3.1$. (d): Graph of $d^2 \gamma^{(QSE)}/d H^2$. The parameters for (a)-(d) are 
$G_0=0.01,\ R_{\gamma_0}=G_1=0.1,\ \rho=10,\ \epsilon_\gamma=\Sigma=0$. 
Values of
$G_0$ and $G_1$ are chosen such that films of all heights are linearly unstable and dewet the substrate when QSE oscillation is not present, that
is, at $\rho=0$;  the absence of QSE oscillation is a hypothetical situation for most metal films. This limit is shown in panel (e) as the contour plot of $\omega\left(k,H_0\right)$.
Notice in this panel, that switching off QSE oscillation completely eliminates the alternation of stable and unstable heights that is seen in panels (a) and (b).
}
\label{FigOmega}
\end{figure}

The non-interacting electron gas model \cite{Han,Wu,OPL} and the DFT calculations \cite{Han} suggest \cite{Hirayama,OPL} the following generic form of $\gamma^{(QSE)}$ in Eq. (\ref{totalEnergy})
for many metal films, including Ag, Pb,  Al,  Na, and Be \cite{Hirayama,Han,Wu}:
\begin{equation}
\gamma^{(QSE)}(h) = \gamma_0^{(QSE)} + \frac{g_0 s^2}{(h+s)^2}\cos{\eta h}-\frac{g_1 s}{h+s},\quad g_0, g_1 > 0,
\label{gammaQSE}
\end{equation}
where the first (constant) term and the second term are due to quantum confinement of the electrons in a thin film, and the third term is due to the
electrons spilling out to the film/substrate interface (the charge spilling effect). 
Eq. (\ref{gammaQSE}) results in the limit of thick film of a \emph{quantum} model, meaning that $N\gg 1$, where $N$ is the number of energy bands occupied by the electrons 
in the $z$-direction. $\gamma_0^{(QSE)} = k_F^2 E_F/(80 \pi) \sim 400-1000$ erg$/$cm$^2$ and $\eta = 2k_F=4\pi/\lambda_F\sim 20$ nm$^{-1}$ for Ag, Au, Mg, Al and Pb. 
(Here $E_F$ is the energy at the bulk Fermi level.)
The energies $g_0$ and $g_1$ also are proportional to $k_F^2 E_F$ (see Ref. \cite{OPL}). 
A small wetting length $s = 0.4 \mbox{nm}\sim 1.2$ ML  (where the estimate of the ML value is based on the atomic diameter of 0.35 nm for Pb) 
prevents divergence as $h\rightarrow 0$ \cite{Chiu}. 
The continuous form (\ref{gammaQSE}) is the result of the interpolation of the set of data points produced by the quantum models \cite{OPL,Han}, whose $z$-coordinates are the 
integer multiples of the monolayer height.
Thus the characteristic amplitudes decays as $\sim 1/h^2$ and $\sim 1/h$ also are the features of the cited quantum models; see Eq. (25) in section 3.3.5, 
and section 3 in Appendix D of Ref. \cite{OPL}, and also section IIC and Fig. 5 in Ref. \cite{Han}. The decay of the dimensionless second derivative of $\gamma^{(QSE)}(h)$ 
with increasing $h$ can be seen in Fig. \ref{FigOmega}(d), and the surface energy itself decays with the same rate. We remark here that QSE oscillation is persistent only in relatively 
thin metal films, as seen in Fig. \ref{FigOmega}(d); Ref. \cite{Ozer} cites the experimental evidence of QSE oscillation in films as thick as 30 ML.

With the choice of $s$ and $\gamma_0$ for the height and energy scales, the dimensionless form of $\gamma^{(QSE)}(h)$ reads:
\begin{equation}
\gamma^{(QSE)}(H) = R_{\gamma_0}+\frac{G_0}{(1+H)^2}\cos{\rho H}-\frac{G_1}{1+H},
\label{gammaQSE2}
\end{equation}
where $R_{\gamma_0}=\gamma_0^{(QSE)}/\gamma_0,\; G_0=g_0/\gamma_0,\; G_1=g_1/\gamma_0$, $H=h/s$ and $\rho=s\eta$. Notice that $0 < R_{\gamma_0},\ G_0,\ G_1 < 1$, and $\rho \sim 10$
for cited values of $s$ and $\eta$.

We assume that the lattice mismatch between the film and the substrate (and more generally, 
the interaction energy of a film and substrate atoms)  has a negligible effect on the
stress at the surface; in other words,
a thin metal film on a substrate is considered free-standing for the purpose of calculating the stress at the surface. Thus the only source of stress in the model is the 
intrinsic surface stress. Clearly, this is justified only for such nearly perfectly lattice-matched quantum growth systems as 
Ag/NiAl(110) \cite{HUQJJLTE}, Pb/Ni(111) \cite{BGZP} and Ag/Fe(100) \cite{MQA} (also see the discussion in Ref. \cite{Han1}). But using large values of the strain parameter (up to the limit of the numerical method) allows the computations to mimic strongly stressed films  due to the lattice mismatch, i.e. Ag/GaAs(110) \cite{Smith,YJEWWNZS}, Ag/Si(111) \cite{Hirayama}, Pb/Si(111) \cite{Ozer,Li} and Pb/Cu(111) \cite{Han1,CHPPM}.
Following the seminal paper by Hamilton \& Wolfer \cite{HW}, in Ref. \cite{JEM_Khenner} we derived the closed-form analytical expression for the surface stress energy.
The dimensionless form of this expression is
\begin{equation}
\gamma^{(S)}(H)=\frac{-\Sigma \Gamma \left[\Gamma + M (H-2)\right]}{\left[2 \Gamma + M (H-2)\right]^2},
\label{gammaSS2}
\end{equation}
where $\Sigma=4 \Gamma P \epsilon_*^2/(\gamma_0 \ell^2)$. 
Also $\Gamma$ ($M$) is the sum of the surface (bulk) elastic moduli, $P$ the total number of atoms in the film, $\epsilon_*$ 
the strain, and $\ell$ the (large) lateral
dimension of the film (the same along the $x$ and $y$ axes).

The anisotropic part of the surface energy (\ref{totalEnergy}) is derived naturally using a thin film scaling \cite{Korzec,Korzec1}. 
Thus the small parameter $\alpha=s/\ell$ is introduced, and then the
dimensionless anisotropic surface energy reads \cite{JEM_Khenner}
\begin{eqnarray}
\gamma^{(Anis)}\left(H_X, H_Y, ...\right) &=& \epsilon_\gamma\left[a_{00} +  \alpha^2\left(a_{20} H_X^2 + \alpha a_{30}  H_X^3 + \alpha^2 a_{40} H_X^4 + a_{02} H_Y^2 + 
      \alpha a_{12} H_X H_Y^2 \right.\right. \nonumber \\ 
&+& \left.\left. \alpha^2 a_{22} H_X^2 H_Y^2 + 
      \alpha^2 a_{04} H_Y^4\right)\right],
\label{gammaAnis2}
\end{eqnarray}
where $\epsilon_\gamma < 0$ is the anisotropy strength, $a_{ij}$'s are the anisotropy coefficients, and $X=x/\ell$, $Y=y/\ell$ are the dimensionless in-plane coordinates.
In this paper we limit the discussion of the anisotropy to [111] surface; for that surface  $a_{00}=1/3$, $a_{20}=4/3$, $a_{30}=-2\sqrt{2}/3$, $a_{40}=-5/2$, $a_{02}=4/3$, $a_{04}=-5/2$, $a_{12}=2\sqrt{2}$, $a_{22}=-5$. \cite{JEM_Khenner}

Finally, the evolution PDE for the time-dependent film surface morphology is derived by first substituting Eqs. (\ref{gammaQSE2}), (\ref{gammaSS2}) and (\ref{gammaAnis2}) 
into Eq. (\ref{totalEnergy}).
Then the total energy $\gamma$ is substituted into the system (\ref{MullinsEq}), (\ref{mu}) - (\ref{mu_hot}), where the adimensionalization is completed using 
the time scaling $t=(\ell^2/D)\hat t$ and the height and in-plane coordinates scalings as shown above. Lastly, a thin film scaling is introduced and the equation is expanded in the powers of $\alpha$, with the dominant 
contributions retained \cite{JEM_Khenner}. This results in the final PDE that is shown in the Appendix.

Primarily for reference, in Fig. \ref{FigOmega}(a-c) the perturbation growth rate $\omega$ from the one-dimensional (1D) linear stability analysis of the PDE is shown for the case of
the isotropic surface energy and zero stress ($\epsilon_\gamma=\Sigma=0$); in the Figure, $k$ is the perturbation wavenumber and $H_0$
the height of the planar film immediately after the low-temperature deposition (the base height). (The expressions for $\omega$ in 1D and 2D cases are listed in Ref. \cite{JEM_Khenner}) From  Fig. \ref{FigOmega}(a,b)  it can be seen that as $H_0$ increases, the planar film surface $z=H_0$ is alternately unstable or stable. 
Such multi-height stability is unusual in a continuum, PDE-based model of a solid film, and it is noted here for the first time.
Larger unstable heights are less unstable
than smaller unstable heights, since the interval of the unstable wavenumbers decreases as $H_0$ increases.
Manifestly, the stable (unstable) $H_0$ values belong to $H$-intervals where  
$\Delta\gamma^{(QSE)}(H)\equiv d^2 \gamma^{(QSE)}/d H^2 > 0$ ($d^2 \gamma^{(QSE)}/d H^2 < 0$), see Fig. \ref{FigOmega}(d) (the thermodynamic stability/instability criterion \cite{Zhang}).
Thus we emphasize here the agreement between the linear stability analysis and the thermodynamic analysis.


%

\section{Nonlinear dynamics of nanoislands}
\label{EvolveIslands}

In this section we numerically study the 1D island dynamics and their transient and equilibrium morphologies. 
We compare the dynamics for the situations when only the QSE is operative; the QSE and anisotropy
are operative; and all three effects, i.e. the QSE, the anisotropy, and the stress are operative.
This way, the understanding of how each physical mechanism contributes to the morphology evolution is obtained. In our model, switching the effect on or off is accomplished simply by taking a non-zero or zero value of the associated governing parameter, 
i.e. $\epsilon_\gamma$ or $\Sigma$.
The 1D setup allows to perform a large number of simulations, and as the result we
present, for the first time in a modeling paper, the complete rationalization of the height and morphology evolution in a small, generic, and idealized quantum system. 
This rationalization is expected to fully carry over to a more realistic 2D system. The parameters set that is chosen for computations is typical. Moderate changes of major parameters
would yield correspondingly moderate, quantitative-only changes of the results, without affecting the overall conclusions and rationalizations. For instance,
slightly changing the amplitudes $G_0$ and $G_1$ of the confinement and spilling terms in the QSE surface energy, Eq. (\ref{gammaQSE2}), would result in the correspondingly moderate 
shifts of the base heights $H_0$, as seen in Fig. \ref{FigOmega}, but the computed island morphologies, heights, volumes, and the coarsening scenarios will not be greatly affected. 

\subsection{The effects of the variation of the base height $H_0$}
\label{EvolveIslands_varyH0}

To elucidate the differences in the morphology evolution at varying $H_0$, here we employ the simple and unique initial condition
in all simulations: two Gaussian-shaped islands on the base surface $z=H_0$, that have the same constant width at half-maximum and the different (but also constant) heights.
Specifically, the initial height (measured from the base surface $z=H_0$) and the width-at-half-maximum 
of the left island are 5 and 8, respectively; these parameters for the right island are 7 and 8.
Thus it can be noticed that the initial island heights 
extend across several stable and unstable height intervals as plotted in Figures \ref{FigOmega}(a,b). 
The horizontal extent of the base surface (along the $X$-axis) is also kept constant at 20$\lambda_{max}$, where $\lambda_{max}=2\pi/k_{max}$ is
the wavelength of the most unstable perturbation from the LSA; here $k_{max}$ is the wavenumber at which the growth rate $\omega$  
attains a maximum value. We used value of $k_{max}$ from Fig. \ref{FigOmega}(c). The base film height $H_0$ is varied in a wide range (from 1.5 to 6). 
For most simulations we chose the linearly stable $H_0$ values listed in the caption to Fig. \ref{FigOmega}.
The boundary conditions for $H(x,t)$ are periodic at $X=0$ and $X=20\lambda_{max}$.

%
\begin{figure}[h]
\centering
\includegraphics[width=6.0in]{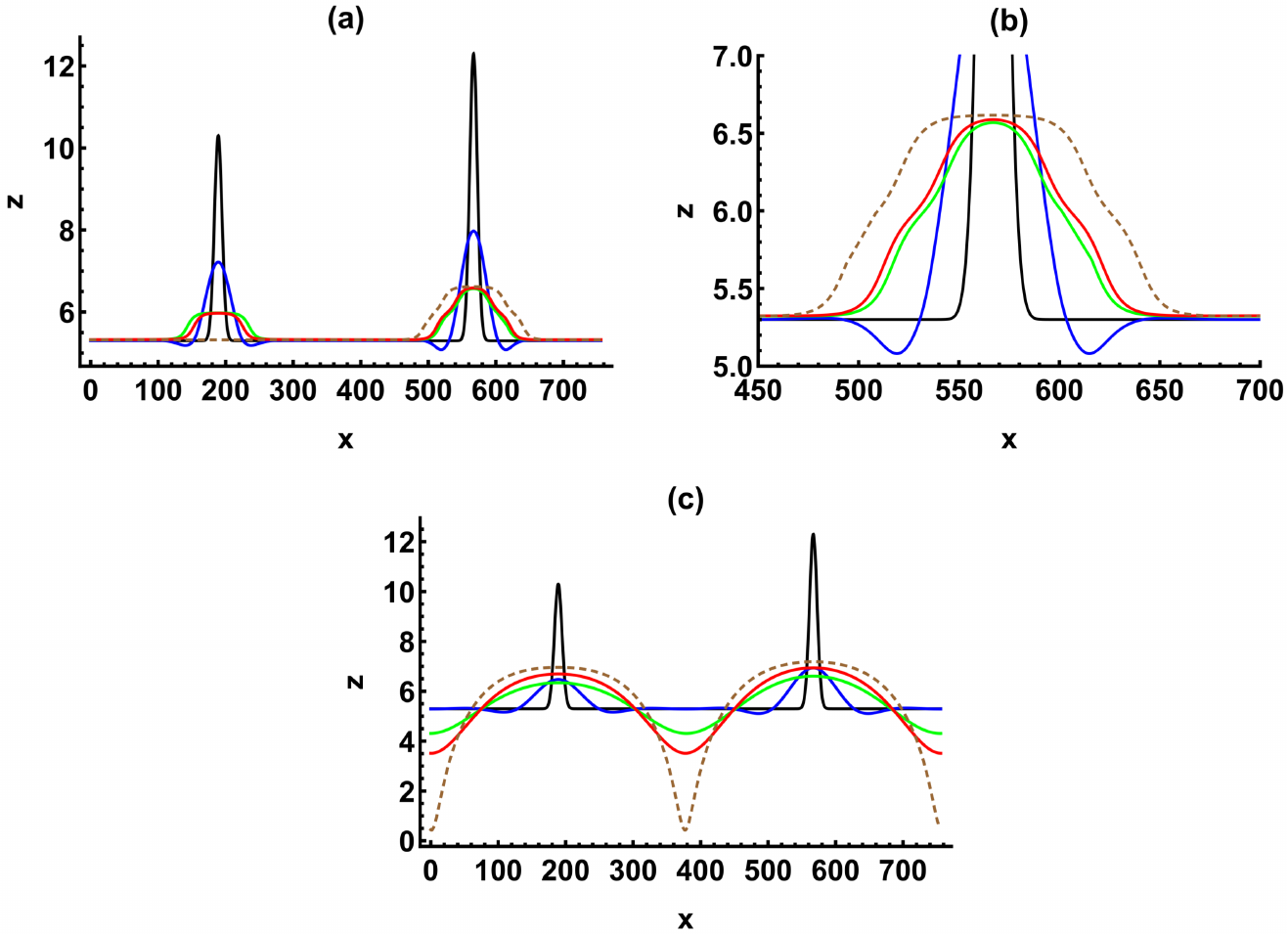}
\caption{(Color online.) (a): Evolution of two islands with the initial Gaussian shape. Dashed line is the final equilibrium morphology, the final time $t_f=10^8$; (b): Magnified view of the right island; 
(c): Islands dewetting. $t_f=1.29\times 10^7$.
$H_0=5.3,\ \epsilon_\gamma=\Gamma=0$, and $\rho=10$ in (a,b), $\rho=0$ in (c). 
}
\label{EvolveIslands_H0=5.3}
\end{figure}

In Fig. \ref{EvolveIslands_H0=5.3} the typical coarsening process can be seen, until the smaller island disappears and the larger one evolves into the equilibrium shape with the flat
top surface. The height of this surface is $H=6.62$, which is one of the stable heights according to Fig. \ref{FigOmega}; the corresponding equilibrium island height
is thus $H_e=6.62-5.3=1.32$, which translates into roughly 1.6 ML (for Pb). Notice that this computation was done without the anisotropy and stress. 
Only for comparison, dewetting of islands is shown in panel (c) at $\rho=0$, i.e. in the case without QSE oscillation.  
Notice that the total initial volume (the one of the bulk film plus the islands) is conserved in Fig. \ref{EvolveIslands_H0=5.3} as well as in other simulations.

When the anisotropy and stress are not present, the typical features of the evolution toward a single equilibrium island seen in Fig. \ref{EvolveIslands_H0=5.3} were also observed in the computations with 
all stable $H_0$ values larger than 2.8. These computations, in particular, show that ``if the lateral size of an unstable island ... is large enough, it will not decay, but rather evolve
to a stable height" \cite{Ozer}; indeed, in Fig. \ref{EvolveIslands_H0=5.3}(a) the larger island develops a larger base when it evolves. However, when $H_0=2.1$ was chosen, the flat top surface disappeared from the island equilibrium shape, i.e. the top became rounded,
and when $H_0=1.5$ was chosen the ``reverse" coarsening was observed - that is, the small island grew at the expense of the large island, and the final equilibrium shape 
was also rounded. Due to the absence of a wetting layer for the chosen parameters (see Fig. \ref{FigOmega}(e)), the atypical reverse coarsening at smaller film heights can be attributed only to a complex nonlinear interaction of the oscillatory QSE terms in the governing PDE (Eq. (\ref{FinalPDE}) in Appendix). We confirmed that, as expected, the dynamics without QSE oscillation 
either in the absence of a wetting layer (Fig. \ref{EvolveIslands_H0=5.3}(c)), or when it is 
present \cite{WettingL} does not show a tendency to reverse coarsening.

\begin{figure}[h]
\centering
\includegraphics[width=3.5in]{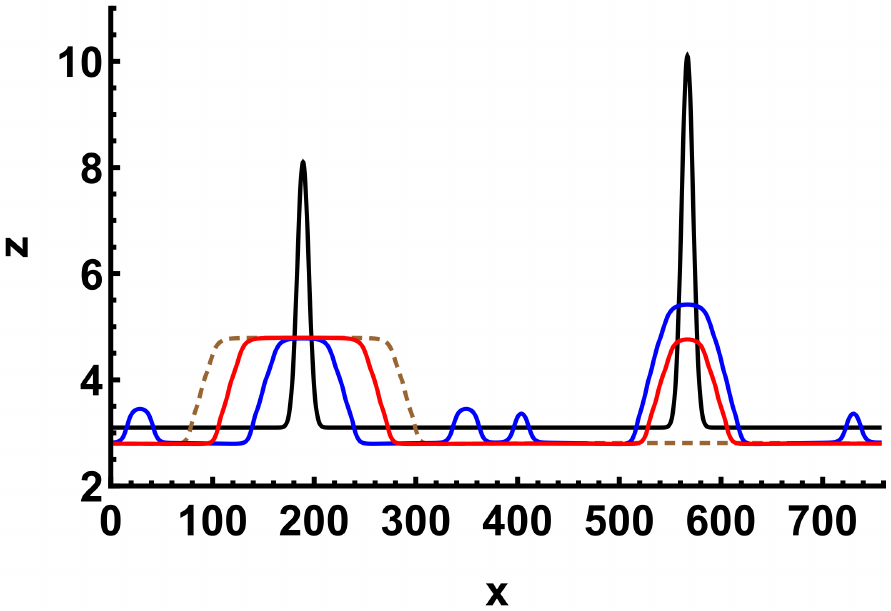}
\caption{(Color online.) Reverse coarsening of two islands for QSE$+$anisotropy$+$stress case at $H_0=3.1$. Dashed line is the island's equilibrium shape.
$\epsilon_\gamma=-0.14,\ \Sigma=0.0053,\ \Gamma=9.6\times 10^{-6},\ M=7.8\times 10^{-6}$, other parameters are as in Fig. \ref{FigOmega}(a-d). $t_f=1.2\times 10^8$. Notice how the left island, after reaching the ``magic height", the large, flat 
and stable top surface $z=4.7$, and with its base now on another stable surface $z=2.8$, starts to grow only laterally, with the constant sidewalls slope; a similar 
evolution can be seen also for the right island in Fig. \ref{EvolveIslands_H0=5.3}(a,b). Such distinct quantum growth dynamics was detected in the STM experiment \cite{UFTE}; it differs qualitatively from the more
familiar Stranski-Krastanow growth mode \cite{Chiu,Korzec,Korzec1,LLRV,Aqua}. Also notice the close resemblance of the computed island shapes to the STM images 
in Fig. 2 of Ref. \cite{Ozer} and in Figures 2 and 17 of Ref. \cite{Han1}.
}
\label{EvolveIslands_H0=3.1}
\end{figure}

When the anisotropy and/or stress effects are turned on, the likelihood of an interrupted coarsening or a reverse  coarsening increases \cite{Korzec,Aqua}. In particular, when the moderate anisotropy is activated, 
the reverse coarsening was found not only at $H_0=1.5$, but also at $H_0=2.1$; with the stress activated, the reverse coarsening is seen at $H_0=2.8$, and the interrupted coarsening
at $H_0=2.1$ and $1.5$; and with both anisotropy and stress activated, the interrupted coarsening is seen at $H_0=1.5, 2.1$ and $2.8$, and the reverse coarsening is seen at 
$H_0=3.1$ and $3.4$. Fig. \ref{EvolveIslands_H0=3.1} shows the reverse coarsening sequence for QSE$+$anisotropy$+$stress case at $H_0=3.1$ (the unstable base surface). 
{\bf Movie 1} (see the \emph{Supplemental Information})
shows the sequence of shape transitions leading to interrupted coarsening for QSE$+$anisotropy$+$stress case at $H_0=2.8$. In the movie, the small island first reaches the stable height
$H=5.3$ (measured from the substrate) and the flat top surface emerges, but because the large island is not stable, the small island increases its height to $H=8$ at the expense of the lateral shrinking; next, the large island decreases its height
until it reaches the stable value $H=4.1$ (and simultaneously the height of the small island increases); the large island then starts to grow again while the small island 
starts getting smaller, etc. These cycles of the slight height growth or incomplete decay seem to repeat, and the equilibrium state of a single island is not reached in a reasonable computation time.

\begin{figure}[h]
\centering
\includegraphics[width=6.0in]{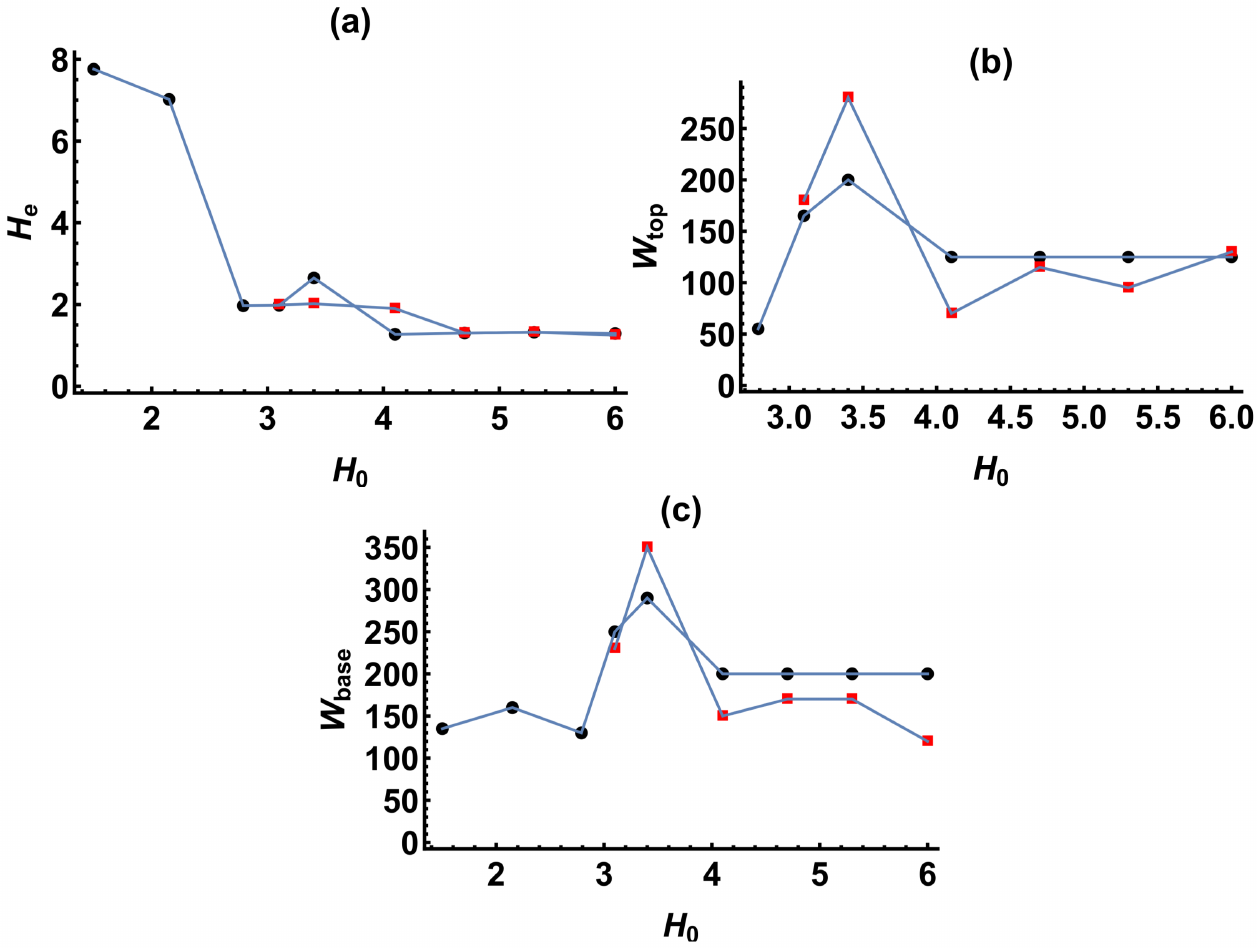}
\caption{(Color online.) (a) The equilibrium island height, (b) width of the top facet, and  (c) width at the base. 
Dots correspond to QSE case, squares correspond to QSE$+$anisotropy$+$stress case with $\epsilon_\gamma=-0.14,\ \Sigma=0.0053,\ \Gamma=9.6\times 10^{-6},\ M=7.8\times 10^{-6}$. The equilibrium island height is measured either from the base surface 
if it is stationary, or from a retracted surface (as in Fig. \ref{EvolveIslands_H0=3.1}); thus this is the ``true" height.
}
\label{IslandHeightWidths}
\end{figure}

For the cases where the single island equilibrium was achieved, Fig. \ref{IslandHeightWidths} shows the equilibrium island height, the width of the top facet, and the width at the 
base as a function of $H_0$.
In the QSE case the height of the island rapidly decreases and reaches the constant value 1.3 ($\sim$1.56 ML for Pb) at $H_0=4.1$; simultaneously, $W_{top}$ and $W_{base}$
also reach constant values 125 and 200, respectively. Notice that the flat top does not emerge until $H_0=2.8$, and then $W_{top}$ first grows then decays. 
From $H_0=2.8$ to $H_0=6$ the dynamics of $W_{base}$ is ``in phase" with $W_{top}$.
In the QSE$+$anisotropy$+$stress case the equilibrium island does not emerge until
$H_0=3.1$ (for this $H_0$ the island is shown in Fig. \ref{EvolveIslands_H0=3.1}), and its height monotonically reaches the same constant value as in the QSE case; the flat top emerges simultaneously with the equilibrium island and its width
tends to the same constant value as in the QSE case, but slower and in the oscillatory ``tail" fashion. A similar trend can be conjectured for $W_{base}$.
Information on the geometric parameters of the islands is not presented in the published experimental papers, but the just discussed predictions of the model may be useful for
future experiments.

Notice that in Fig. \ref{EvolveIslands_H0=3.1} the base surface recedes from its initial unstable height at $H_0=3.1$ to a nearest lower stable height $H=2.8$; 
this also happens without the anisotropy and the stress. (Similar decomposition of an initial unstable blanket coverage into a stable blanket coverage and stable islands on top
is observed in the experiment \cite{CHPPM}.) Obviously, in the computation such surface retraction does not prevent the islands from evolving and coarsening.
This is attributed to either of the two factors.  First, if a small deformation at the base of the island emerges on a horizontal length scale which is 
much smaller than $\lambda_{max}$, then such short-wavelength perturbation quickly decays since $\omega<0$ for small perturbation wavelengths, see Fig. \ref{FigOmega}(c).    
Second, if such deformation is of an unstable wavelength, then it stops growing ($\omega$ becomes zero) as soon as its amplitude reaches 
the stable film height immediately below $H_0$. In this case the film surface will fairly quickly recede to this new stable height, as seen in Fig. \ref{EvolveIslands_H0=3.1}.
Since deformations of a second kind are characteristic of a nonlinear island dynamics even when $H_0$ is stable, the surface retraction was noted for several such 
$H_0$ values ($H_0 = 1.5, 2.1, 2.8$ and 3.4). {\bf Movie 2} (see the \emph{Supplemental Information}) shows the details of the retraction process for $H_0=3.4$; in a process similar 
to the coarsening sequence shown in Fig. \ref{EvolveIslands_H0=5.3}, the smaller left island
in the movie later is absorbed into the right island. 
At larger $H_0$ the retraction is arrested, which seems to indicate that the 
QSE again plays a role. The small satellite islands that are seen on the receded surface (Fig. \ref{EvolveIslands_H0=3.1}) on a path to equilibrium are another hallmark of the two-islands dynamics on initial unstable base surfaces. 
In fact, in the beginning of the coarsening the size of these islands may be only slightly less than the size of either left or right island. For comparison, such satellite islands
do not emerge on the receding stable base surface, as seen in the {\bf Movie 2}.


\subsection{The effects of the variation of the island volume}

In this section we study the height transitions of the equilibrium island as a function of the increasing island volume, at fixed $H_0$ (and only for QSE case). 
Real islands have various volumes at different stages on a path to equilibrium; to our knowledge, the island height transitions due to volume variations were not systematically reported in the experimental publications.
We chose $H_0=5.3$ to eliminate the base surface retraction; for the example see Fig. \ref{EvolveIslands_H0=5.3}.
The initial condition is again two Gaussian-shaped islands, which mimics the real-life situation - since in the experiment there is always multiple islands or none.
The initial height (measured from the base surface $z=5.3$) and width-at-half-maximum 
of the left island are 1.3 and 8, respectively, and the height of the right island is 3.2; its width at half maximum is increased from 8 to produce the
increasing volume $V$.
The smaller left island disappears fast, and the equilibrium height of the right island is plotted vs. 
$V$.

\begin{figure}[h]
\centering
\includegraphics[width=3.5in]{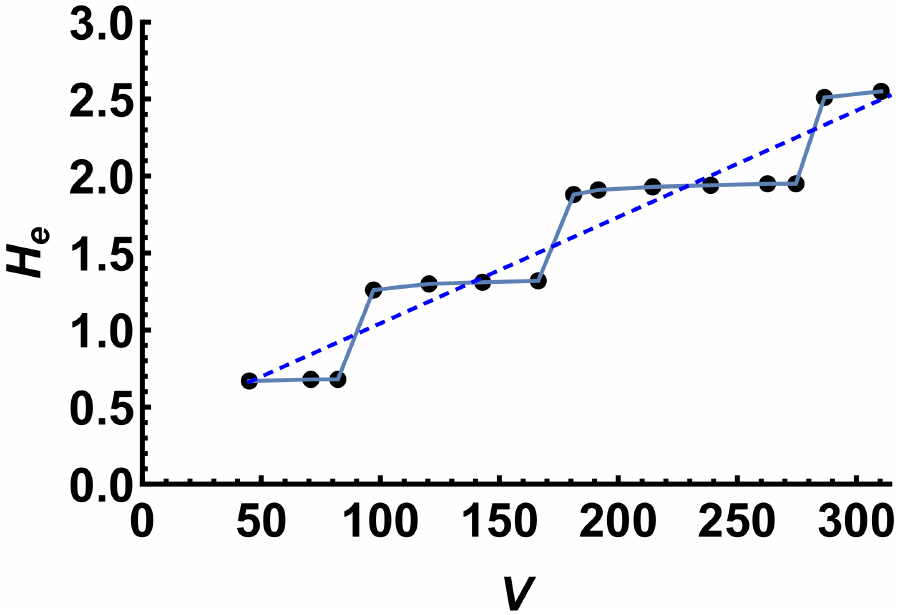}
\caption{(Color online.) Equilibrium island height vs. its volume. $H_0=5.3$, the parameters are as in Fig. \ref{FigOmega}(a-d). The equation of the linear fit
(dashed line) is $H_e = 0.007 V + 0.35$. $H_e$ is measured from the base stationary surface $z=5.3$.
}
\label{He_vs_V}
\end{figure}

From Fig. \ref{He_vs_V} it can be concluded that the new equilibrium island height, if measured from the substrate, is selected from the set of stable heights between 6 and 7.8 (see the caption to Fig. \ref{FigOmega}) every time $V$ reaches a new critical value; notice that the transitions are sharp. It takes 83\% volume growth to increase $H_e$ from 6 to 6.6 (or from 0.7 to 1.3, as in the Figure); 71\% volume growth to increase $H_e$ from 6.6 to 7.2 (from 1.3 to 1.9); and 51\% volume growth to increase $H_e$ from 7.2 to 7.8 (from 1.9 to 2.5).

The 
QSE again exerts a strong influence through value of the film height $H_0$. Decreasing $H_0$ from 5.3 to 1.5 while keeping all other parameters constant 
(including the initial heights of both islands) resulted in the 
data points arrangement that is visually similar to Fig. \ref{He_vs_V}, however, it takes only 60\% volume growth to increase $H_e$ from 2.8 to 3.4 (if measured from the substrate; or from 1.3 to 1.9, if measured from the base surface); 26\% volume growth to increase $H_e$ from 3.4 to 4.1 (from 1.9 to 2.6); and 36\% volume growth to increase $H_e$ from 4.1 to 4.7 (from 2.6 to 3.2). 
The slope of the linear fit to the data points for this case is 0.012, and thus the conclusion is that at smaller $H_0$ the height of the equilibrium island increases faster with the increase of the island volume. 

\subsection{Coarsening of a pitted surface into the ``magic" islands}

In this section we discuss the results of the computations of the surface coarsening, assuming that the starting base surface is ``pockmarked" by irregularly shaped pits.
The depth of all pits is chosen randomly, such that in a given run some pits initially may be shallow and other pits may be deep. 
Also, the pits number density along the base surface is a constant for all runs, but their positioning and shapes are random.
This is a rather common initial condition in the experiments \cite{YJEWWNZS,MQA}. Our primary goal is to understand qualitatively the major factors that govern the outcomes of the coarsening process. We will show that there is two such factors: 
first, the stability or instability of the base surface, and second, the presence of deep pits.

\begin{figure}[h]
\centering
\includegraphics[width=6.0in]{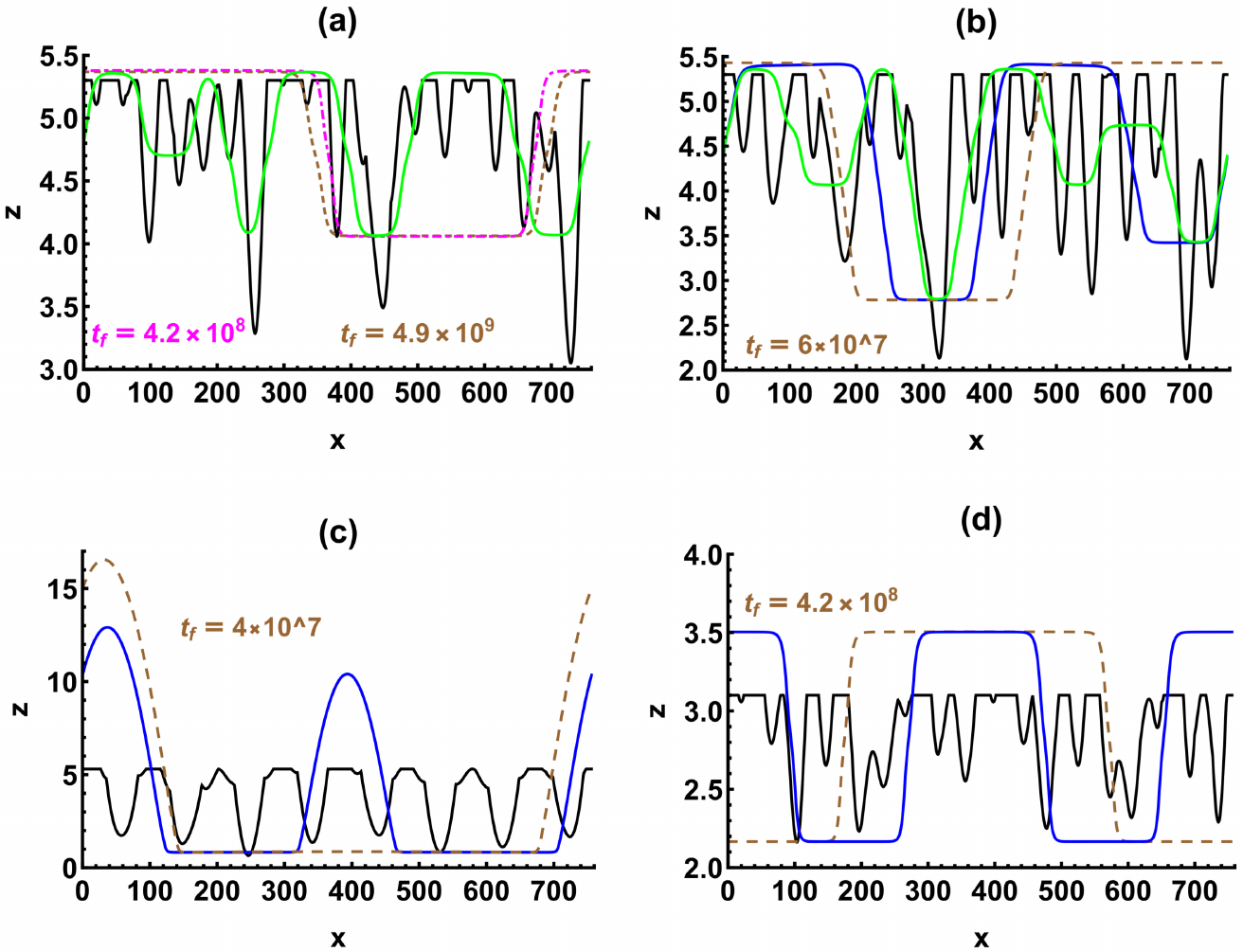}
\caption{(Color online.) (a-d): Coarsening of a very rough surface. Dashed lines show one-island equilibrium surface morphologies attained at the dimensionless final time marked
in each panel. These computations were done at zero anisotropy and stress, i.e. at the parameters of Fig. \ref{FigOmega}(a-d). 
Dash-dotted line in panel (a) is the final equilibrium morphology for the
QSE$+$anisotropy$+$stress case with $\epsilon_\gamma=-0.14,\ \Sigma=0.0053,\ \Gamma=9.6\times 10^{-6},\ M=7.8\times 10^{-6}$, starting from the same initial base surface.
Solid lines show the initial and a few intermediate morphologies.
}
\label{Random_IniCond_Island_Coarsening}
\end{figure}

In Fig. \ref{Random_IniCond_Island_Coarsening} we collected the results of four representative computations. In the panels (a)-(c) the base surface is stable at $H_0=5.3$, 
and a few deepest pits are of increasing depth from (a) to (c). 
In the panel (a), where the pits are shallow, the coarsening was computed until two large flat-top islands were formed. They ``sit" on the stable surface $H=4.1$ and
their top surface is at the stable level $H=5.4$. Thus the islands extend vertically across one stable domain with the center at $H=4.7$.  When the depth of the deepest
pit increased slightly from $d=2.3$ (panel (a)) to $d=3.3$ in panel (b), the coarsening rate sharply increased - notice that the single flat-top island formed by $t=6\times 10^7$. 
The height of this island also reacted to the increase of the pit depth by stabilizing its base at a lower stable height $H=2.8$, making the island extend across three stable domains with 
the centers at $H=3.4, 4.1, 4.7$.
When the pits are even deeper ($d=4.3$) 
as in panel (c), the outcome of the coarsening changes dramatically. The islands now form on a stable surface
whose position roughly coincides with the tip of the deepest pit, and most important, the flat-top ``magic" islands do not emerge. Instead a fairly narrow and very high islands with a curved top and sidewalls emerge and coarsen.

In panel (d) the base surface is at the unstable height $H_0=3.1$, and the pits are shallow. 
The flat-top islands are formed
similar to the scenarios seen in panels (a) and (b). As in panel (a), two flat-top islands 
extend vertically across one stable domain. Characteristically, the island's top surface is at the nearest (to $H_0=3.1$) larger stable height $H=3.5$ (despite that another, smaller 
stable height $H=2.8$ is also available; thus it is easier to increase the island height rather than to decrease it and increase the width). 

Lastly, the anisotropy and the stress accelerate coarsening; indeed, the single large island shown by the dashed line in panel (a) has been formed
by $t_f=4.2\times 10^8$, vs. $t_f=4.9\times 10^9$ for the QSE case. However, these factors do not affect the height of the island and the vertical positioning of its base and top surfaces.

\begin{figure}[h]
\centering
\includegraphics[width=6.0in]{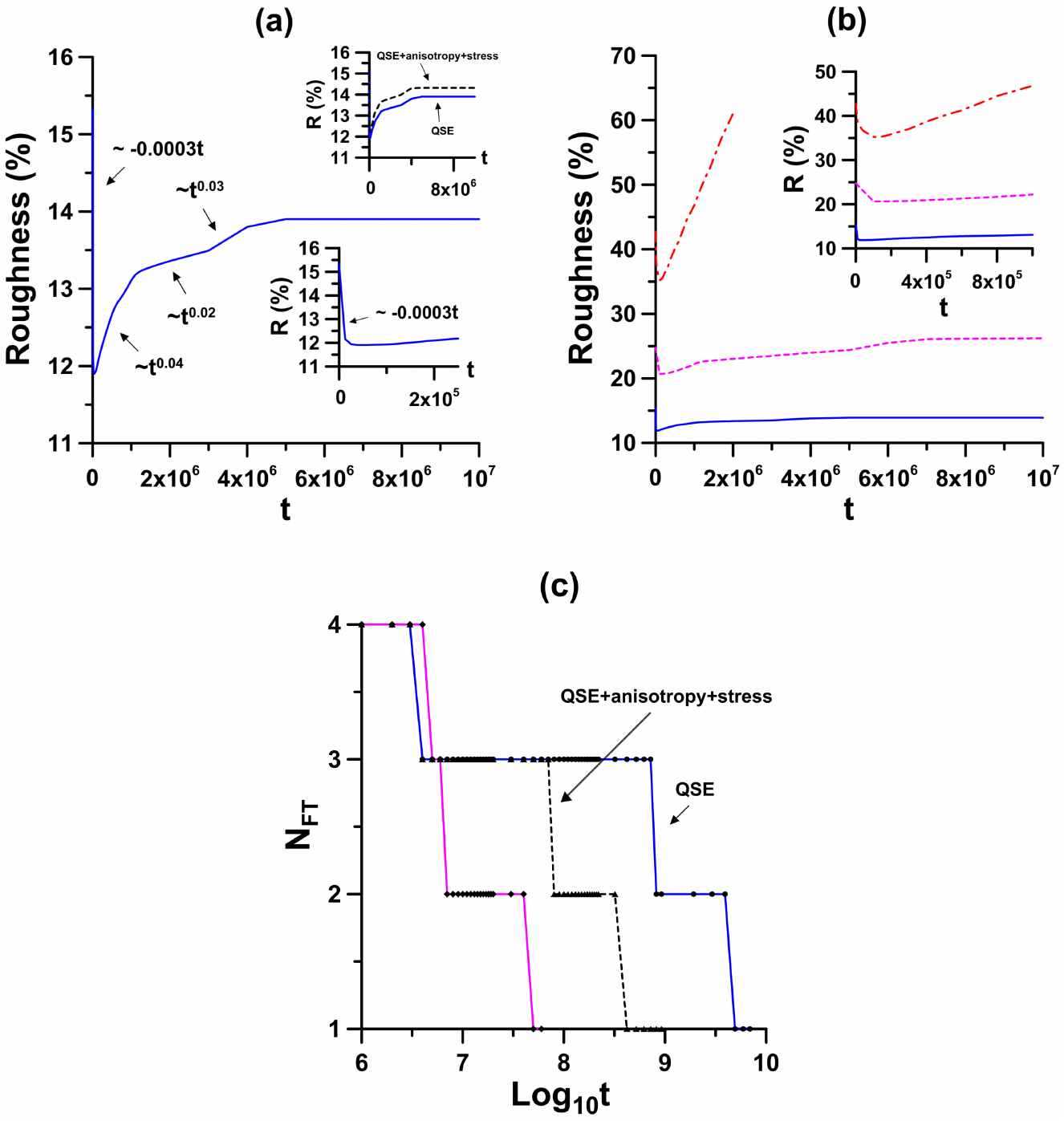}
\caption{(Color online.) (a,b): Surface roughness vs. time. Solid line in panels (a) and (b) corresponds to evolution without anisotropy and stress shown in Fig. 
\ref{Random_IniCond_Island_Coarsening}(a). Dashed and dash-dotted lines in panel (b) correspond to evolution (also without anisotropy and stress) shown in Fig. 
\ref{Random_IniCond_Island_Coarsening}(b,c), respectively; thus this panel compares the roughness from Figs. \ref{Random_IniCond_Island_Coarsening}(a-c), while 
panel (a) shows the roughness from Fig. \ref{Random_IniCond_Island_Coarsening}(a) separately. 
The insets show the zooms at small times; the inset in panel (a) compares the cases with and without anisotropy and stress,
as in Fig. \ref{Random_IniCond_Island_Coarsening}(a). (c): The number, $N_{FT}$, of ``magic" flat-top islands vs. time. The lines are only the guides for the eye.
The circles (diamonds) correspond to evolution without anisotropy and stress shown in Fig. \ref{Random_IniCond_Island_Coarsening}(a) (Fig. \ref{Random_IniCond_Island_Coarsening}(b)).
The triangles correspond to evolution with anisotropy and stress shown in Fig. \ref{Random_IniCond_Island_Coarsening}(a).
}
\label{Random_IniCond_Measures}
\end{figure}

In Fig. \ref{Random_IniCond_Measures} the quantitative measures of coarsening in Figs. \ref{Random_IniCond_Island_Coarsening}(a-c) are provided. It can be seen that in these three cases 
the root-mean-square roughness first decreases fast (linearly in time) for a very short period, resulting in the formation of pre-``magic" islands with the curved sidewalls and
the top surface, as shown by the green line in Figs. \ref{Random_IniCond_Island_Coarsening}(a,b). For the cases of initially shallow and moderately deep pits
(Figs. \ref{Random_IniCond_Island_Coarsening}(a,b)) this is followed by the power laws dynamics until roughness reaches a constant value, which coincides with the 
emergence of four flat-top ``magic" islands (Fig. \ref{Random_IniCond_Measures}(c)). Further coarsening of these islands into the single equilibrium island 
shown in Figs. \ref{Random_IniCond_Island_Coarsening}(a,b) proceeds with islands not changing their heights, thus the roughness is unchanged. 
This scenario is contrasted to the case of initially deep pits in Fig. \ref{Random_IniCond_Island_Coarsening}(c). In that case, after the initial decrease, the roughness grows linearly in time (Fig. \ref{Random_IniCond_Measures}(b)).
This growth is capped only by the termination of the coarsening, i.e. when a single, large ``non-magic" island emerges, as seen in Fig. \ref{Random_IniCond_Island_Coarsening}(c).

The results in this section point out again that the pits in the film surface serve as the initiation sites for the morphological transition, strongly 
mediated by the QSE, toward the flat-top islands \cite{YJEWWNZS,MQA}. Also we remark that these results and the results in Sec. \ref{EvolveIslands_varyH0} re-emphasize and cast some new light on the importance of the QSE strength (through the film height) for the formation and dynamics 
of the metal nanoislands \cite{Ozer,Li,Han1}.


\section{Conclusions}

A model that incorporates the QSE, anisotropy, and surfaces stress has been used to compute evolution to equilibrium of 
the pre-existing nanoscale metal islands and the coarsening of a severely rough (pitted) surface. 
Some of the most common qualitative 
features of the morphological evolution of metal nanoislands 
reported in the experiments on various metal/substrate systems are reproduced. The kinetic pathways in a formation and coarsening of the quantum islands are clarified.

We showed that the nonlinear dynamics of the morphology evolution is complicated and that it depends on factors such as 
the initial (base) film height $H_0$ after the deposition, the levels of anisotropy and stress and, for a pitted surface, whether the pits 
are deep.
The dependence on the base height is not only on a height value itself, which sets the strength of the QSE,
but also whether $H_0$ belongs to one 
of the stable or unstable height intervals imposed by the QSE. In most situations the surface evolves into the ``magic" islands that have the shape of a truncated pyramid 
with a wide, flat top surface and a well-defined height, which is a difference of two stable base heights (again as imposed by the QSE). Interrupted 
coarsening
is often found at a small base height, and/or active anisotropy and strong stress. 
The dependence of the geometric parameters of the equilibrium island on the base height and on the island initial volume was computed. The equilibrium island height,
width of the base, and width of the flat top surface reach constant values as the base height increases. The gradual increase of the island initial volume at a constant base height leads to the abrupt increases of the equilibrium island height. 

It should be again noted that we employed approximations or assumptions in order to obtain a concise and numerically tractable model of the complicated multi-physics phenomena.
They are: a small aspect ratio of the film (height/lateral dimension), small lattice mismatch, and 1D modeling. 

The first assumption holds with overwhelming accuracy in the thin film systems such as discussed in the paper, since the film height and the lateral dimension typically are of the order of 10 ML and 10-100 microns, respectively. Our mathematical treatment merely formalizes this vast difference of the spatial scales by expanding the extended Mullins equation (\ref{MullinsEq}) (after the adimensionalization) in powers of a small parameter, which is the above ratio, and retaining the dominant contributions. The result is the PDE that contains all physics of the original equation, but unlike the original equation, it allows not only the direct computation,
but also the stability analysis of the equilibria. As we show in the paper, that analysis, when it goes hand-in-hand with computation, is instrumental and indispensable 
for understanding the formation of ``magic" islands (see the discussions of Figures \ref{FigOmega}, \ref{EvolveIslands_H0=3.1} and \ref{Random_IniCond_Island_Coarsening}). \cite{FTN}

In regard to strain/stress effects, we chose in this paper to limit the consideration to the intrinsic surface stress. This is because this effect is important and was cited as the primary
source of stress in many systems of interest \cite{HUQJJLTE,BGZP,MQA} and also, because a theoretical treatment so far has been insufficient.
Our model of the surface stress paves way to continuum models of surface stress-dominated morphology evolution of thin films (either metal or semiconductor).  
In a comprehensive model that is presently under development, this treatment is augmented by the traditional model of the lattice mismatch stress \cite{SDV,GolovinPRB2004,Korzec1}, thus both sources of 
stress are considered and their combined effects on nanoscale metal islands can be computed. Since QSE strongly dominates and defines (at large) how such islands form and evolve,
the augmentation by the lattice mismatch stress is expected to only provide a numerical correction and not affect the qualitative features of the dynamics of islands' morphology.

Lastly, 1D modeling is useful to distill the key features of the 3D model, extract the dynamics that is at least semi-quantitatively correct, and perform a parametric
analysis in order to explore a range of experimental conditions, such as the varying levels of stress and anisotropy (as is done in this paper). Fine features of the islands' shape
and the corrections to coarsening rates will be in future computed with the help of a 3D model, which is unavoidably computationally intensive. Thus only a single set of the material parameters for a particular metal/substrate system can be employed by such model.

By accounting for the lattice mismatch stress, growth, and the
complicated, non-trivial QSE oscillation (which is unique to a chosen metal/substrate combination), the comprehensive model will allow computation of the morphology evolution
and coarsening of several tens of 3D metal nanoislands. 
We expect that this will further advance our understanding of these technologically important systems and help design the film growth protocols that would result in better height, 
shape and position selectivity of the nanoislands.

\section{Appendix}

The 1D version of the governing PDE reads
\begin{eqnarray}
H_t &=& B\left[\frac{G_1}{1+H}\left(H_{\text{XXXX}}-\frac{2}{(1+H)^2}H_{\text{XX}}-\frac{2}{1+H}H_XH_{\text{XXX}} -\frac{\left((1+H) H_{\text{XX}}-2H_X^2\right)H_{\text{XX}} }{(1+H)^2}+\frac{6 }{(1+H)^3}H_X^2\right)\right. \nonumber\\
&-&\left(1+R_{\text{$\gamma_0$}}\right)H_{\text{XXXX}}+\frac{G_0}{(1+H)^2}\left(\frac{6-\rho ^2(1+H)^2}{(1+H)^2}H_{\text{XX}}-H_{\text{XXXX}}+\frac{4 }{1+H}H_XH_{\text{XXX}}\right.\nonumber\\
&-&\left.\frac{\left(6 H_X^2
-2(1+H)H_{\text{XX}} -(1+H)^2H_X^2 \rho ^2 \right)H_{\text{XX}}}{(1+H)^2}+\frac{\left( 6(1+H)^2\rho ^2-24 \right)H_X^2}{(1+H)^3}\right)
\cos{\rho  H}\nonumber\\
&+&\frac{G_0\rho }{(1+H)^2}\left(\frac{4}{1+H}H_{\text{XX}}+2 H_XH_{\text{XXX}}-\frac{\left(4H_X^2 -(1+H) H_{\text{XX}}\right)H_{\text{XX}}}{1+H}+\frac{\left((1+H)^2\rho
^2-18 \right)H_X^2}{(1+H)^2}\right) \sin{\rho  H}\nonumber\\
&-&\epsilon _{\gamma }\left(a_{00}+2a_{20}\right)H_{\text{XXXX}}-4\epsilon _{\gamma }a_{20}\left(2H_{\text{XXX}}^3+6H_XH_{\text{XX}}H_{\text{XXX}}+H_X^2H_{\text{XXXX}}\right)\nonumber\\
&+&\frac{ \Sigma \Gamma  (\Gamma  +(H-2) M)}{(2 \Gamma  +(H-2) M)^2}\left( H_{\text{XXXX}} -\frac{2M}{2 \Gamma  +(H-2) M}H_{\text{XX}}^2-\frac{4M }{2\Gamma  +(H-2) M} H_XH_{\text{XXX}}\right.\nonumber\\
&-&\left.\frac{6M^2 }{(2 \Gamma  +(H-2) M)^2}H_X^2H_{\text{XX}}\right)+\frac{3 \Sigma \Gamma  (H-2)M^3}{(2 \Gamma  +(H-2) M)^4}\left(\frac{4M}{2 \Gamma  +(H-2) M}H_X^2- H_{\text{XX}} \right)\nonumber\\
&+&\frac{ \Sigma \Gamma M}{(2 \Gamma  +(H-2) M)^2}\left(H_{\text{XX}}^2+2H_XH_{\text{XXX}}-\frac{6 M^2}{(2 \Gamma  +(H-2) M)^2}H_X^2 +\frac{M}{2
\Gamma  +(H-2) M}H_{\text{XX}}\right.\nonumber\\
&-&\left.\left.\frac{4M}{2\Gamma  +(H-2) M}H_X^2H_{\text{XX}} \right)\right].
\label{FinalPDE}
\end{eqnarray}
Here $B=\Omega^2 \nu \gamma_0/\ell^2 \alpha kT$ is Mullins number; 
in the computations
$B$ is set equal to one. Also, $t$ stands for the dimensionless time (that is, the hat over $\hat t$ is dropped).
The terms proportional to $\cos{\rho H}$ and $\sin{\rho H}$ originate in the oscillatory contribution to the surface energy due to QSE (Eq. \ref{gammaQSE2}). Also we remark that
in the typical case $\Gamma > M$ there is no singularity of the stress terms, as $2\Gamma + (H-2)M>0$ at all $H\ge 0$.


\begin{thebibliography}{200}
\bibitem{WettingL} A wetting layer can be introduced by increasing $G_0$ while keeping $G_1=0.1$; say, $G_0=1/15$ results in a wetting layer of the dimensionless height one.
This follows from the expression for the critical film height \cite{JEM_Khenner},  $H_{0c}=3G_0/G_1-1$, above which all heights are linearly unstable and below which all heights are stable.
At the chosen value $G_0=0.01$ for Fig. \ref{FigOmega} and other figures in the paper, $H_{0c}<0$ and thus all film heights are unstable - as shown in Fig. \ref{FigOmega}(e).
\bibitem{FTN} This approach, which bears
the names longwave expansion, thin film scaling, or lubrication approximation has been fruitfully used in the theoretical/modeling thin film community for decades; see Refs. \cite{SDV,WD} for some of the first applications to solid and liquid films, respectively.
\end{thebibliography}
\end{document}